\documentclass{elsart}
\usepackage{amssymb}
\usepackage{epsfig}

\begin{document}

\begin{frontmatter}

\title{Search for exotic baryons in double radiative capture 
on pionic hydrogen.}

\author{P.A.~{\.Z}o{\l}nierczuk\thanksref{UK}}$^{\spadesuit}$
\author{D.S.~Armstrong\thanksref{WM}}  
\author{E.~Christy\thanksref{UK}}$^{\heartsuit}$ 
\author{J.H.D.~Clark\thanksref{WM}}$^{\diamondsuit}$
\author{T.P.~Gorringe\thanksref{UK}} 
\author{M.D.~Hasinoff\thanksref{UBC}}
\author{M.A.~Kovash\thanksref{UK}} 
\author{S.~Tripathi\thanksref{UK}} 
\author{D.H.~Wright\thanksref{TRIUMF}}$^{\clubsuit}$

\address[WM]{College of William and Mary, Williamsburg, VA 23187}
\address[UK]{University of Kentucky, Lexington, KY 40506}
\address[UBC]{University of British Columbia, Vancouver, B.C., Canada V6T 1Z1}
\address[TRIUMF]{TRIUMF, 4004 Wesbrook Mall, Vancouver, B.C., Canada V6T 2A3} 

\begin{abstract}
We report a search 
for low-lying exotic baryons via double radiative capture
on pionic hydrogen. 
The data were collected
at the TRIUMF cyclotron
using the RMC spectrometer
by detecting gamma-ray pairs
from pion stops in liquid hydrogen.
No evidence was found to support an earlier claim for exotic baryons 
of masses 1004 and 1044~MeV/$c^2$.
We obtain upper limits
on the branching ratios
for double radiative capture
via these exotic states
of $< 3 \times 10^{-6}$ 
and $< 4 \times  10^{-6}$ 
respectively.
\end{abstract}

\begin{keyword} 
\end{keyword}

\end{frontmatter}

\section{Introduction}


Until recently all hadrons were classified 
as either three quark 
or quark-antiquark 
states. The mounting evidence
\cite{Nakano03,Barmin03,Stepanyan03,Barth03,Aratayn03,Kubarovsky03}
for an exotic baryon 
-- the $\theta^+ (1540)$ --
has likely begun a new chapter in hadron spectroscopy.
The $\theta^+ (1540)$ has baryon number $B = +1$ 
and strangeness $S = +1$,
which implies a minimally exotic pentaquark substructure $uudd\bar{s}$.
Such exotics would provide
a new venue for exploring the rich dynamics of QCD
and testing the multifarious models of hadrons. 


In this light the recent claim 
by Tatischeff {\it et al.}\ \cite{Tatischeff97,Tatischeff02}
of evidence for low-lying exotic baryons 
between the nucleon and the $\Delta$ resonance 
is quite intriguing. 
Tatischeff {\it et al.}\
investigated the $p p \rightarrow p \pi^+ X$ reaction 
using the SPES3 spectrometer at the Saturne Synchrotron
by detecting the forward-going $\pi^+$/$p$ pairs 
from $p$-$p$ collisions at energies of $1520$-$2100$~MeV.
From the $\pi^+$/$p$ momenta they reconstructed
the missing mass $M_X$ for $pp \rightarrow p \pi^+ X$ events 
and found several small peaks.
The observed intensities of the peaks were about $10^{3}$
times smaller than 
the intensity of the neutron peak from the 
$p p \rightarrow p \pi^+ n$ reaction.
They observed two peaks below the $\pi$N threshold
corresponding to masses of $1004$~MeV/$c^2$ and $1044$~MeV/$c^2$,
with widths of $<15$~MeV/$c^2$.
The authors have claimed these structures as evidence 
of low-lying exotic baryons.


Their claim has rekindled  
an old suggestion by Azimov 
of a low-lying exotic baryon octet \cite{Azimov70,Azimov03}.
The catalyst for the original work was the difficulty 
of the multiplet assignment for the 
$\Sigma( 1480)$.
Azimov suggested identifying 
the strangeness $S = -1$ $\Sigma (1480)$ \cite{Pan70,Engelen80},
the strangeness $S = -1$ $\Lambda (1330)$ \cite{Ross72,Briefel77},
and strangeness $S = -2$ $\Xi (1620)$ \cite{Bogachev69},
as three members of a new low-lying exotic baryon octet.\footnote{We 
caution the reader that the $\Lambda (1330)$ is not listed,
and the  $\Sigma( 1480)$ and $\Xi (1620)$ have only one-star status,
in the Particle Data Group listings \cite{Hagiwara02}.}
Azimov then used the Gell-Mann-Okubo mass formula 
to estimate the mass 
of the lowest-lying non-strange member of the multiplet 
to be roughly $1100$~MeV/$c^2$.
We shall follow the convention adopted in
Refs.\ \cite{Azimov70,Azimov03} and denote this isodoublet by $n^{\prime}$.
More recently in Ref.\ \cite{Azimov03}  the authors have 
considered the possibility 
of a pentaquark substructure
for this exotic octet,
and discussed the relation between the octet
containing the $n^{\prime}$ 
and the anti-decuplet 
containing the $\theta^+ (1540)$.
They argue that $J^{\pi} = 1/2^-$
is the most likely spin-parity
for the $n^{\prime}$ state.

Other attempts to explain the 
baryon candidates of Tatischeff {\it et al.}\ 
are also available.
These include such ideas as a
fully antisymmetric $q^3$ spin-flavor state
\cite{Kobushkin98},
colored $q\bar{q}-q^3$ and $qq-q$ quark clusters
\cite{Tatischeff02,Noya02},
and the collective excitations of the
$q\bar{q}$ condensate
\cite{Walcher01}.
We refer the reader to Ref.\ \cite{Tatischeff02}
for a fuller discussion of these ideas.

Unfortunately the claims for exotic baryons by Tatischeff {\it et al.}\ 
\cite{Tatischeff97,Tatischeff02}
were not corroborated in two subsequent electro-production experiments.
Jiang {\it et al.}\ \cite{Jiang03} employed
the two high resolution spectrometers in Hall A at JLab  
to search for evidence of narrow baryon resonances 
in the $e p \rightarrow e \pi^+ X$ reaction
at an incident energy of $1.7$~GeV.
They found no evidence of peaks 
in the $e p \rightarrow  e \pi^+ X$ missing-mass spectrum,
and set upper limits of $7 \times 10^{-4}$ 
for the $1004$~MeV/$c^2$ state and  $8 \times 10^{-4}$
for the $1044$~MeV/$c^2$ state, with respect
to  neutrons from $e p \rightarrow e \pi^+ n$.
Likewise Kohl {\it et al.}\ \cite{Kohl03}, 
who employed the three-spectrometer facility at MAMI
to study the  $e p \rightarrow  e \pi^+ X$ reaction at $855$~MeV,
found no evidence for any peaks in the $e p \rightarrow  e \pi^+ X$
missing-mass spectrum in the mass range $970$-$1060$~MeV/c$^2$.
They established an upper limit on electro-production 
of exotic baryons of about $10^{-4}$ the yield 
of the neutrons from the $e p \rightarrow  e \pi^+ n$ reaction.


The null results 
from the electro-production experiments
do not necessarily conflict
with the hadro-production experiment
since a hadro-production channel for exotic baryons
may be absent in the electro-production case \cite{Jiang03}.
Further, the different reactions and the different kinematics
may have unique sensitivities to such exotic baryons,
thus preventing the direct comparison of hadro-production
yields and electro-production limits.


Another concern,
first raised by  L'vov and Workman \cite{Lvov98}
and later discussed by Azimov {\it et al.}\ \cite{Azimov03},
is that low-lying exotic baryons would contribute 
to Compton scattering on protons and neutrons.
More specifically, the $1004$~MeV/$c^2$ state 
would produce a peak at $68$~MeV and the $1044$~MeV/$c^2$ state
would produce a peak at $112$~MeV 
in nucleon Compton scattering.
However the cross section is proportional 
to the radiative width of the  $n^{\prime} \rightarrow n \gamma$ transition,
thus ultra-narrow baryons could escape detection in Compton scattering.
Indeed, using the available Compton scattering data,
the authors of Refs.\ \cite{Azimov03,Lvov98} 
have set limits of roughly $1$-$10$~eV on the 
radiative widths of the exotic baryon candidates
from the Tatischeff {\it et al.}\ experiment.


In Refs.\ \cite{Azimov70,Azimov03}
the authors have further observed
that exotic baryons below the pion-nucleon threshold
could additionally contribute 
to double radiative capture on pionic hydrogen atoms,
{\it i.e.} $\pi^- p \rightarrow \gamma \gamma n$.
Specifically, the $n^{\prime}$ could contribute 
through a two-step sequence involving
$n^{\prime}$ production 
and subsequent radiative decay
\begin{eqnarray}
\pi^- p  & \rightarrow & n^{\prime} \gamma \\ 
\nonumber
& & ~\hookrightarrow n \gamma .
\end{eqnarray}
Indeed Azimov {\it et al.}\ 
have used the measured branching ratios for single
and double radiative capture 
to set a conservative upper limit 
on $n^{\prime}$ production 
of $B.R. ( \pi^- p \rightarrow n^{\prime} \gamma )
/ B.R. ( \pi^- p \rightarrow n \gamma )
 < 8 \times 10^{-5}$ \cite{Azimov03}.

\begin{figure}
\begin{center} 
\mbox{\epsfig{figure=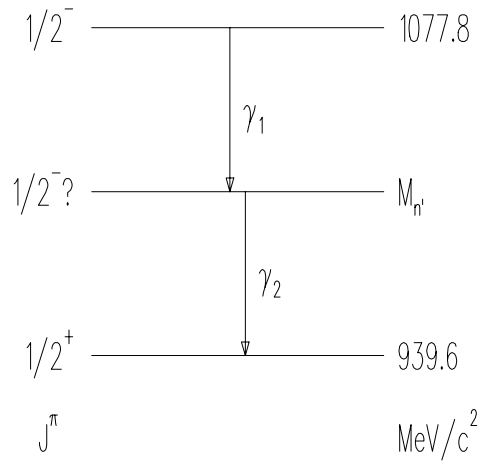,height=6.0cm}}
\end{center}
\caption{The $n^{\prime}$ mediated contribution to
double radiative capture on pionic hydrogen.
The diagram shows the spin-parities and masses
for the relevant states in the gamma-ray cascade
$ \pi^- p \rightarrow n^{\prime} \gamma  \rightarrow n \gamma \gamma$.} 
\label{fig1}
\end{figure}

The energetics of 
double radiative capture via $n^{\prime}$ production 
are shown in Fig.\ \ref{fig1}.
The production mechanism yields a mono-energetic gamma-ray 
and the $n^{\prime}$ decay produces a Doppler broadened gamma-ray.
The gamma-rays energies are therefore determined by the $n^{\prime}$ mass,
with $E_{\gamma 1} = 71$~MeV and $E_{\gamma 2} \simeq66$~MeV
for a 1004~MeV/$c^2$ $n^{\prime}$ state, 
and $E_{\gamma 1} = 33$~MeV and $E_{\gamma 2} \simeq 99$~MeV
for a 1044~MeV/$c^2$ $n^{\prime}$ state.
The Doppler width $2 E_{\gamma 2} ( v / c )$ 
of the decay $\gamma$-ray 
is about 9.3~MeV for $M = 1004$~MeV/$c^2$, and 
about 6.3~MeV for $M = 1044$~MeV/$c^2$.
Additionally, assuming the $n^{\prime}$ baryon has 
spin-parity $J^{\pi} = 1/2^-$ \cite{Azimov03},
the two-step process in Fig.\ \ref{fig1} involves a spin-parity sequence 
of $1/2^- \rightarrow 1/2^- \rightarrow 1/2^+$
and yields an isotropic angular correlation
between the production and decay gamma-rays.

In this article we report a search for evidence 
of exotic baryons with masses below the $\pi$N 
threshold via double radiative capture
on pionic hydrogen.
The experimental data were originally collected
to study the direct two-photon process
of double radiative capture,
and are reported in Tripathi {\it et al.} \cite{Tripathi02}.
Herein we have re-analyzed the data 
to search for evidence of any exotic baryons
in the mass range $970$-$1050$~MeV/$c^2$.
In Secs.\ \ref{expt} and \ref{backgrounds} we briefly describe 
the experimental setup and background sources
for the measurement.
In Sec.\ \ref{events}  we discuss 
our approach in searching for 
any $n^{\prime}$-mediated events 
and in Sec.\ \ref{limits} we discuss 
our method for setting limits on $n^{\prime}$-mediated capture.
We summarize in Sec.\ \ref{conclusions}.


\section{Experimental setup}
\label{expt}

The experiment was conducted at the TRIUMF cyclotron using the 
RMC spectrometer.
Incoming pions 
of flux $7 \times 10^{5}$~s$^{-1}$
and momentum 81.5~MeV/$c$
were counted in a 4-element plastic scintillator telescope
and stopped in a 3~liter liquid hydrogen target.
Outgoing photons were detected 
by pair production 
in a 1~mm thick cylindrical lead converter 
and e$^+$e$^-$ tracking in cylindrical multi-wire and drift chambers.  
A 1.2~kG axial magnetic field was used for momentum analysis
and concentric rings of segmented scintillators 
were used for fast triggering.
The trigger scintillators comprised: 
an A-ring located inside the lead converter, 
a C-ring located between the lead converter and the tracking chambers,
and a D-ring located outside the drift chamber.
For more details on the RMC spectrometer
see Wright {\it et al.}\ \cite{Wright92}.


We employed
a two--photon trigger 
that was based
on the multiplicities 
and the topologies
of the hits   
in the trigger scintillators
and the drift chamber cells.
The A-ring scintillators were used to reject events
that had charged particles originating from the target region.
The pattern of C and D-ring scintillators hits and drift cells hits
were used to select events with 
topologies resembling the two $e^+$e$^-$ pairs
from two-photon events \cite{Tripathi02}.


During a four week running period
we recorded $\pi^- p \rightarrow \gamma \gamma n$ data
from $4.0 \times 10^{11}$ pion stops
in liquid hydrogen.
Calibration data 
with a dedicated $\pi^o \rightarrow \gamma \gamma$ trigger
were also collected.


\section{Background sources}
\label{backgrounds}

The signature of $n^{\prime}$ production 
in double radiative capture is a coincident $\gamma$-$\gamma$ pair 
with one mono-energetic gamma-ray
and one nearly mono-energetic gamma-ray
with an energy sum $E_{\gamma 1} + E_{\gamma 2} \simeq m_{\pi}$
(ignoring the $n^{\prime}$ recoil energy and the neutron-proton mass 
difference).

The $n^{\prime}$ mechanism will compete 
with direct two-photon production
from double radiative capture
on pionic hydrogen.
The later process yields a photon pair with individual energies 
$E_{\gamma 1}$, $E_{\gamma 2}$
and summed energy $E_{\gamma 1} + E_{\gamma 1}$ 
that span that full 
$\pi^- p \rightarrow \gamma \gamma n$ three-body phase space.
The branching ratio for double radiative capture on pionic hydrogen
has been measured by Tripathi {\it et al.}\ \cite{Tripathi02}
and calculated by Joseph \cite{Joseph60}, 
Lapidus and Musakhanov \cite{Lapidus72}, 
and Beder \cite{Beder79}.
The measurement by Tripathi {\it et al.}\ yielded
$B.R. = ( 3.09 \pm 0.44 ) \times 10^{-5}$
and the tree-level calculation by Beder
yielded $B.R. = 5.1  \times 10^{-5}$,
in fair agreement.
The photon energy and angle distributions
from experiment and theory 
are in good agreement.
Thus, together the small branching ratio of the direct two-photon process
and distinctive kinematics of the $n^{\prime}$ mediated process
are sufficient to separate these two sources of photon pairs.

One source of two-photon background 
is real $\gamma$--$\gamma$ coincidences
from $\pi^o \rightarrow \gamma \gamma$ decay.
The $\pi^o$'s were produced by
pion charge exchange $\pi^- p \rightarrow \pi^0 n$,
both at-rest and in-flight.
The branching ratio for at-rest charge exchange
is $B.R. = 0.6$ and yields $\pi^o$'s with energies $T = 2.8$~MeV 
and photon pairs with opening angles $\cos{\theta} < -0.91$.
The much smaller contribution from in-flight charge exchange
yields $\pi^o$'s with energies $T < 15$~MeV and photon
pairs with opening angles  $\cos{\theta} < -0.76$.
Consequently the $\pi^o$ background overwhelms
other two-photon signals at large opening angles.
Thus the direct two-photon process and 
the $n^{\prime}$ mediated process
are only detectable at smaller angles.


Another source of background
was accidental $\gamma$--$\gamma$ coincidences
arising from multiple $\pi^-$ stops.
The pion beam 
had a micro--structure with a pulse width of 2--4~ns,
and a pulse separation of 43~ns.
With an incident flux of $7 \times 10^{5}$~s$^{-1}$,
the probability for more than one pion arriving in a single beam pulse
is about 1.5\%.
Multiple pion stops in one beam pulse
can yield a $\gamma$-$\gamma$ pair
by the accidental coincidence of one photon
from the first pion and another photon from the second pion.
These accidental $\gamma$-$\gamma$ coincidences 
yield photon-pairs with opening angles $-1.0 < \cos{\theta} < +1.0$
and summed energies 106-258~MeV.

In subsequent discussions we shall denote the 
real $\gamma$-$\gamma$ coincidences from $\pi^o$ decay
the ``$\pi^o$ background'', and the 
accidental $\gamma$-$\gamma$ coincidences from 
multiple pion stops the ``2$\pi$ background''.

\section{Event selection}
\label{events}

The data analysis was a three step process.
In step one we applied cuts to identify photon pairs 
that originated from the target.
In step two we applied cuts to reject the contributions
from the $\pi^o$ background and the 2$\pi$ background
and select the events from either double radiative capture 
or $n^{\prime}$ mediated capture.
In step three we used the distinctive photon energy distributions
of double radiative capture 
and $n^{\prime}$ mediated capture
to search for evidence of the later process.

To identify photon pairs we applied both a tracking cut
and a photon cut.
The tracking cut imposed minimum values for the number of points 
in the drift chamber tracks,
and maximum values for the chi--squared of the fits.
The photon cut 
required that the electron-positron pairs intersect at the lead converter
and that the reconstructed photon pairs originate from the hydrogen target.
A total of $2.3 \times 10^6$ photon pairs were found to survive 
these cuts.
These events were dominated by the $\pi^o$
and 2$\pi$ backgrounds.

To reject the  $\pi^o$ and 2$\pi$ backgrounds
we applied a beam counter amplitude cut, a C-counter timing cut,
and a two-photon opening angle cut.
The beam counter amplitude cut rejected events with
large energy deposition in the beam scintillators,
thus indicating the arrival of multiple pions in one beam pulse.
The C-counter timing cut rejected events with 
large time differences between the C-counter hits, 
thus indicating the two photons to have originated 
from two neighboring beam pulses.
The opening angle cut eliminated events with
$\cos{ \theta } < -0.1$, rejecting
the background from $\pi^o \rightarrow \gamma \gamma$ decay.
A total of 635 events with summed energies 
$ > 80$~MeV was found  to survive all cuts.

\begin{figure}
\begin{center} 
\mbox{\epsfig{figure=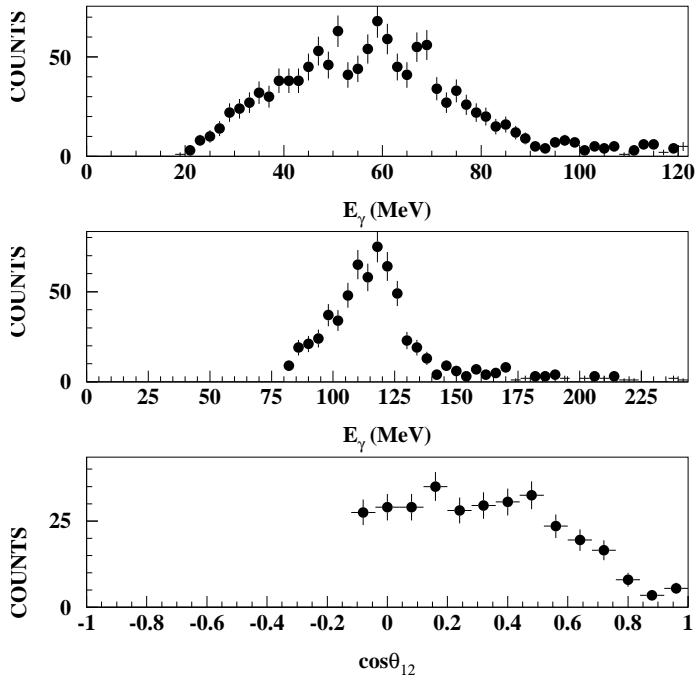,height=10.0cm}}
\end{center}
\caption{The individual photon energy spectrum (top), 
summed photon energy spectrum (center), 
and two-photon opening angle spectrum (bottom),
for the 635 events that passed all cuts.
Note the individual energy spectrum contains $2 \times 635$
photons.} 
\label{fig2}
\end{figure}

The individual energy, summed energy, and opening angle
spectra for the 635 events that survived all cuts
are shown in  Fig.\ \ref{fig2}. 
The summed energy spectrum shows a peak
at  $E \sim m_{\pi}$.\footnote{The peak's centroid is
shifted downward due to energy loss of the $e^+$$e^-$ pairs
in traversing the lead converter, trigger scintillators, etc.} 
The individual energy spectrum 
shows a broad continuum
with a low energy cut-off at about 20~MeV 
and a high energy cut-off at about 100~MeV
due to the acceptance of the spectrometer.
The opening-angle spectrum shows the 
minimum two-photon opening angle
at $\cos\theta  = -0.1$.

Note that the spectra in Fig.\ \ref{fig2} contain a
small residual contribution 
from the $\pi^o$ background
and the 2$\pi$ background.
These contaminations were estimated in Ref.\ \cite{Tripathi02}
at $( 8.3 \pm 1.1 )$\% 
and $( 6.0 \pm 0.9 )$\%, respectively.

\section{Baryon sensitivity}
\label{limits}

A maximum likelihood procedure \cite{Barlow93}
was employed to search for evidence of any 
$n^{\prime}$-mediated events.
Specifically, we fit the individual photon energy spectrum 
in Fig.\ \ref{fig2} to the function 
\begin{equation}
f( E_{\gamma} ) = 
N_{n^{\prime}} f_{n^{\prime}}( E_{\gamma} ) +
N_{d} f_d( E_{\gamma} )
\end{equation}
where $f_{n^{\prime}} (E_{\gamma} )$ 
and $f_d (E_{\gamma})$ are distributions that describe
the expected energy spectra of $n^{\prime}$-mediated capture 
and direct two-photon capture,
and $N_{n^{\prime}}$ and $N_d$ are the corresponding amplitudes
of the two processes.

To obtain the distributions $f_{n^{\prime}} (E_{\gamma} )$ 
and  $f_d (E_{\gamma} )$,
which involve a convolution of their
true energy distributions
with the detector response function,
we used a Monte Carlo simulation.
The Monte Carlo was based on the CERN GEANT3 package \cite{GEANT}
and incorporated  both the detailed geometry of the RMC detector
and the detailed interactions of the various particles.
The simulation was tested by comparison to photon spectra
from at-rest pion radiative capture $\pi^- p \rightarrow \gamma n$ and
at-rest pion charge exchange $\pi^- p \rightarrow \pi^o n$.
For further details see Refs.\ \cite{Tripathi02,Zolnierczuk02}.

For the $n^{\prime}$-mediated process the photon energy spectrum
has a double-peak structure with the production gamma-ray centered at
approximately 
$E_{\gamma 1} \simeq ( M_{\pi p} - M_{n^\prime} ) = 
( 1078 - M_{n^\prime} )$~MeV
and the decay gamma-ray centered at approximately 
$E_{\gamma 2} \simeq ( M_{n^\prime} - M_{n} ) = ( M_{n^\prime} - 940 ) $~MeV.
However, in the mass region $1000$~MeV/c$^2$ $< M_{n^\prime} <$ 1020~MeV/c$^2$,
these two peaks are overlapping because of the finite resolution
of the pair spectrometer and the Doppler broadening
of the decay gamma-ray.
In addition, for either $M_{n^{\prime}} < 970 $ ~MeV/c$^2$ 
or $M_{n^{\prime}} > 1050$ ~MeV/c$^2$, 
the photon pairs becomes unobservable because 
of the low energy cut-off of the pair spectrometer.

For the direct two-photon process the photon energy spectrum 
is a three-body continuum. The spectrum was calculated
by Beder in Ref.\ \cite{Beder79} and measured 
by Tripathi {\it et al.}\ in Ref.\ \cite{Tripathi02}.
Given the agreement between the calculation
and the measurement,
we employed the Beder calculation
of the $\gamma$-energy spectrum shape
for simulating the direct two-photon process.
The resulting energy spectrum, 
after the convolution with the response function
of the pair spectrometer,
is a broad continuum with a
low energy cut-off at about 20~MeV,
a high energy cut-off at about 100~MeV,
and a maximum at about $60$~MeV.

\begin{figure}
\begin{center} 
\mbox{\epsfig{figure=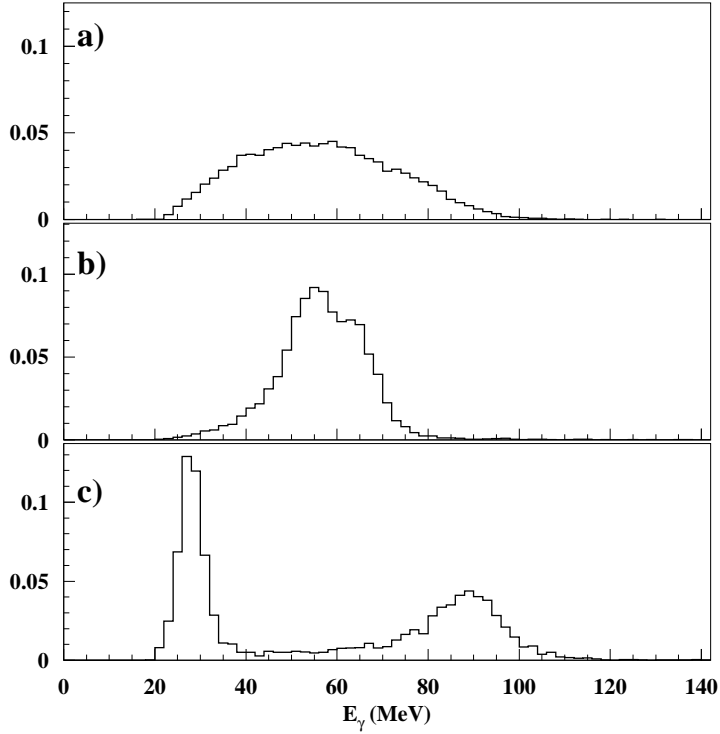,height=10.0cm}}
\end{center}
\caption{The functions $f_d ( E_{\gamma} )$ (top),
$f_{n^{\prime}} (E_{\gamma} )$ with $M_{n^{\prime}} = 1004$~MeV/c$^2$ (center),
and $f_{n^{\prime}} (E_{\gamma} )$ with $M_{n^{\prime}} = 1044$~MeV/c$^2$ (bottom),
that were utilized in the fitting procedure.
The functions were obtained by Monte Carlo simulation
in order to convolute the detector response function
and theoretical line-shape.} 
\label{fig3}
\end{figure}

The resulting distributions for describing
the direct two-photon process 
and $n^{\prime}$-mediated process
are shown in Fig.\ \ref{fig3}.
For $f_{n^{\prime}} (E_{\gamma} )$
we plot the distributions
for both $M_{n^{\prime}} = 1004$~MeV/c$^2$
and $M_{n^{\prime}} = 1044$~MeV/c$^2$.
In the latter case the production gamma-ray
and decay gamma-ray are well separated
and the double-peaked $f_{n^{\prime}} (E_{\gamma} )$
and broad continuum $f_d ( E_{\gamma} )$
are easily distinguished.
However in the former case the
production gamma-ray and decay gamma-ray are overlapping,
and the discrimination between the line-shapes
is reliant on the different widths 
of the two distributions
(the full width half maximum
is about $25$~MeV for the $n^{\prime}$-mediated process,
and about $50$~MeV for the direct two-photon process).
Consequently, our sensitivity to $n^{\prime}$ production
is dependent on the mass of the $n^{\prime}$ state.

In the maximum likelihood procedure
we stepped the value of $M_{n^{\prime}}$
from $970$~MeV/c$^2$ to $1050$~MeV/c$^2$ in increments of $10$~MeV/$c^2$
and varied the ratio of $n^{\prime}$ mediated events to
direct two-photon events, {\it i.e.} $N_{n^{\prime}}$/$N_d$.
Note that (i) the fit was restricted to $20 < E_{\gamma} < 100$~MeV
in order to avoid the 2$\pi$ background events at energies above $100$~MeV,
and (ii) we forced the sum $N_{n^{\prime}} + N_d$ 
to equal the total number of photon pairs observed in the 
region of the fit.
The procedure yielded values
of $N_{n^{\prime}}$/$N_d$ that were consistent 
with zero for all values of $M_{n^{\prime}}$.
Thus no evidence for $n^{\prime}$-mediated capture was found 
for $970 < M_{n^{\prime}} < 1050$~MeV/c$^2$,
the measured spectrum being completely consistent 
with direct two-photon capture only.


In the absence of any $n^{\prime}$ signal, we employed the above procedure 
to establish the 90\% confidence limit 
on the $N_{n^{\prime}}$ counts
as a function of the mass $M_{n^{\prime}}$.
To convert the  upper limit on the counts to an upper limit
on the branching ratio for the $n^{\prime}$-mechanism 
we used 
\begin{equation}
\label{e:br}
B.R. = \frac{N_{n^{\prime}}} { N_{\pi^-} \cdot \epsilon\Omega \cdot F \cdot C} 
\end{equation}
where $N_{\pi^-}$ is the number 
of live-time corrected pion stops,
and $\epsilon\Omega \cdot F$ is the detector acceptance.
The acceptance was obtained 
using the Monte Carlo
for $n^{\prime}$ production.
The factor $F = 0.90 \pm 0.09$ \cite{Tripathi02} accounts 
for detector inefficiencies
which are present in the experiment 
but are absent in the simulation.
The factor $C = 0.85 \pm 0.01$ \cite{Wright98} 
accounts for the fraction of the incoming pions 
that stop in the hydrogen target.

\begin{figure}
\begin{center} 
\mbox{\epsfig{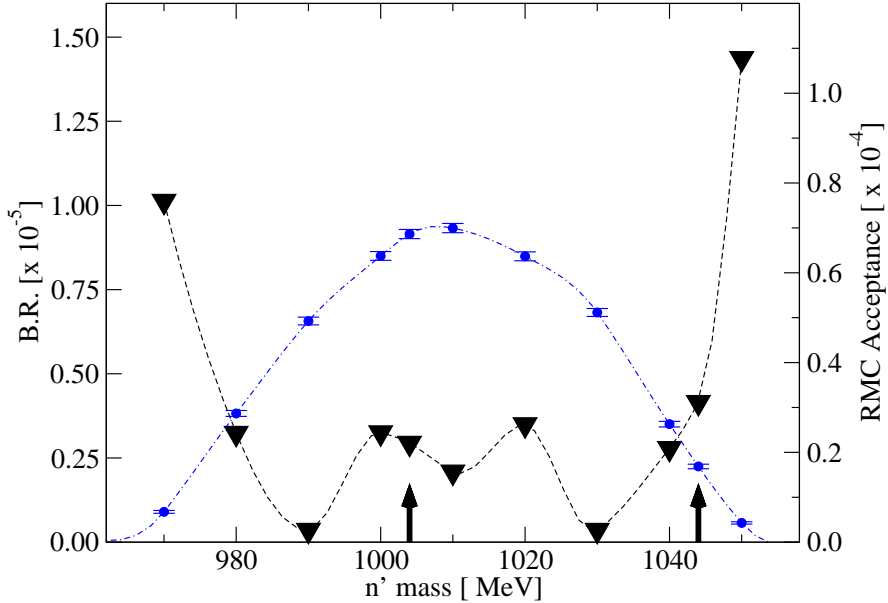}}
\end{center}
\caption{
The 90\% confidence limit (solid triangles) 
for the branching ratio  
of $n^{\prime}$ production 
via double radiative capture
as a function of the $n^{\prime}$ mass.
Also shown, via the filled circles and the right-hand scale,
is the two-photon acceptance versus the $n^{\prime}$ mass
from the Monte Carlo simulation.
Note the dashed line (branching ratio curve)
and dot-dashed line (two-photon acceptance curve) are 
merely to guide the eye.} 
\label{fig4}
\end{figure}

The resulting branching ratio limit for $n^{\prime}$ production 
in $\pi^-$p capture is shown in Fig.\ \ref{fig4}.\footnote{These limits 
assume an isotropic two-photon angular distribution
for the $n^{\prime}$ contribution to the double radiative capture reaction.}
The double-well shape of the branching ratio limit 
is a consequence of two factors. 
Firstly, for either low or high masses, the
$n^{\prime}$ detection efficiency decreases sharply 
and hence the $n^{\prime}$ branching ratio limit increases sharply.
Secondly, for masses from 1000 to 1020~MeV/c$^2$, the 
overlapping production and decay gamma-rays
make discrimination between the direct two-photon
and $n^{\prime}$-mediated capture more troublesome 
and hence the $n^{\prime}$ branching ratio limit
also rises. For $M_{n^{\prime}} = 1004$~MeV/c$^2$
we obtained the 90\% C.L. limit $BR < 3 \times 10^{-6}$ 
and for  $M_{n^{\prime}} = 1044$~MeV/c$^2$
we obtained $BR < 4 \times 10^{-6}$.

In Ref.\ \cite{Azimov03} the authors have employed the
$n^{\prime}$ mechanism in double radiative capture
to establish a limit on the $n^{\prime} \rightarrow n \gamma$ 
radiative width $\Gamma_{\gamma} ( n^{\prime} )$. 
Their scheme for estimating  $\Gamma_{\gamma} ( n^{\prime} )$
was based on the suppression between the radiative capture
to the $n^{\prime}$ state, {\it i.e.} $\pi^- p \rightarrow \gamma n^{\prime}$,
and the neutron state, {\it i.e.} $\pi^- p \rightarrow \gamma n$.
The authors of Ref.\ \cite{Azimov03} 
have used this suppression factor to estimate 
the $n^{\prime}$ radiative width from the $\Delta$ radiative width 
by assuming the same factor also relates the
$n^{\prime} \rightarrow n \gamma$ decay 
and $\Delta \rightarrow n \gamma$ decay.
Of course this method is rough and ready -
for example, in neglecting the different multi-polarities
for the radiative decays of the $n^{\prime}$ state 
($1/2^- ? \rightarrow 1/2^+$) 
and the $\Delta$ resonance
($3/2^+ \rightarrow 1/2^+$).
However, while noting this reservation, by applying the method
of Ref.\ \cite{Azimov03} we obtained limits 
for $n^{\prime}$ radiative decay
of $\Gamma_{\gamma} ( n^{\prime} ) <  0.05$~eV for 
the $M_{n^{\prime}} = 1004$~MeV/c$^2$ candidate
and $\Gamma_{\gamma} ( n^{\prime} ) <  0.3$~eV for 
the $M_{n^{\prime}} = 1044$~MeV/c$^2$ candidate.
For comparison, the limits
from Compton scattering 
are $\Gamma_{\gamma} ( n^{\prime} ) <  0.2$~eV 
and $\Gamma_{\gamma} ( n^{\prime} ) < 1.6$~eV 
respectively \cite{Lvov98}.


\section{Summary}
\label{conclusions}

Using the RMC spectrometer at the TRIUMF cyclotron we 
have searched for contributions of low-lying exotic
baryons to double radiative capture on pionic hydrogen.
We found no evidence for exotic baryons of masses 1004 
and 1044~MeV/$c^2$ as claimed by Tatischeff {\it et al.}\ 
\cite{Tatischeff97,Tatischeff02},
and  set upper limits
on the branching ratios
of double radiative capture 
via these exotic states
of $< 3 \times 10^{-6}$ 
and $< 4 \times  10^{-6}$ 
respectively.
Our result, together with null results  
from pion electro-production \cite{Jiang03,Kohl03}
and nucleon Compton scattering \cite{Azimov03,Lvov98},
are important constraints on any models
of such baryons.
Moreover the absence of confirmation
of Tatischeff's claims 
in subsequent experiments
now casts some doubt on the existence of such low-lying states.
A further hadro-production experiment
is required to settle the issue.


The authors are indebted to Profs.\ 
Yakov Azimov and Igor Strakovsky for prompting our efforts
and many valuable discussions.
We wish to thank the staff 
of the TRIUMF laboratory
for their support of this work.
In particular we acknowledge the help
of Dr.\ Ren\'{e}e Poutissou on the data acquisition,
and Dr.\ Dennis Healey on the hydrogen target.
In addition we thank the National Science Foundation (United States),
the Natural Sciences and Engineering Research Council (Canada), 
the Clare Boothe Luce Foundation (JHDC), 
and the Jeffress Memorial Trust (DSA),
for financial support.

\end{document}